\documentclass[12pt]{article}
\usepackage[T1]{fontenc}
\usepackage[latin1]{inputenc}

\makeatletter

\providecommand{\LyX}{L\kern-.1667em\lower.25em\hbox{Y}\kern-.125emX\@}

 \newcommand{\lyxaddress}[1]{
   \par {\raggedright #1 
   \vspace{1.4em}
   \noindent\par}
 }

\usepackage {graphicx}
\usepackage {longtable}

\makeatother

\begin{document}

\title{Monte Carlo simulation of abnormal grain growth in two dimensions }

\author{Ren\'e Messina\( ^{a,} \)\thanks{
Now at : Max Planck Institut f\"ur Polymerforschung, Ackermannweg 10, 55128
Mainz, Germany \protect \\
email: messina@mpip-mainz.mpg.de 
}, Mich\`{e}le Soucail\( ^{a} \) and Ladislas Kubin\( ^{b} \)}

\maketitle

\lyxaddress{\( ^{a} \)ONERA, DMMP, 29 Av. de la Division Leclerc, BP 72, 92322 Chatillon
Cedex, France }

\lyxaddress{\( ^{b} \)LEM, CNRS-ONERA, 29 Av. de la Division Leclerc, BP 72, 92322 Chatillon
Cedex, France }


\begin{abstract}
Abnormal grain growth in the presence of second phase particles is investigated
with the help of a two-dimensional Monte Carlo simulation. An aggregate of equiaxed
grains is considered with constant grain boundary energy and mobility. The only
driving force accounted for stems from the grain boundary curvature. The process
of abnormal grain growth is investigated as a function of two governing parameters,
the initial degree of pinning of the matrix grains by the particles and the
initial size advantage of the anomalous grain. In such conditions, moderate
growth is obtained whose specific features are discussed with respect to the
available models. It is shown that it is possible to obtain drastic grain growth
by introducing the thermally activated unpinning of grain boundaries from particles.
For this purpose, a simplified but effective procedure is proposed and discussed
that includes the influence of the capillary force on the height of the local
energy barrier for grain unpinning. 
\end{abstract}

\section{Introduction}

The grain growth processes occurring in polycrystalline materials submitted
to an annealing treatment are usually classified into two types. The first type
corresponds to a self-similar coarsening process and is called normal grain
growth. It is characterised by a uniform increase of the grain size with a time-invariant
distribution function. The second type is called abnormal grain growth and is
characterised by the coarsening of a few grains at the expense of the surrounding
matrix. In such conditions, the self-similar aspect of the coarsening process
is lost. In addition, one also distinguishes a particular kind of abnormal grain
growth, which is sometimes called \char`\"{}drastic\char`\"{} abnormal grain
growth. In this last case, a few grains can invade the whole microstructure
during an annealing treatment. The final size of these grains can be two orders
of magnitude larger than the grain size of the surrounding matrix grains and,
in some cases, a single crystal is eventually obtained. Such heterogeneous microstructures
may be highly detrimental to mechanical properties and it is particularly important
to avoid their occurrence during the thermo-mechanical processing of fine-grained
materials. 

Different mechanisms have been proposed to explain abnormal grain growth. A
small prestrain can promote the onset of anomalous grain growth by a mechanism
of critical strain annealing. In this case, a critical deformation is needed
for the growth of a few grains that will consume all the others, the driving
force being the energy stored in the deformed grains by dislocation networks.
This mechanism has been reproduced with the help of Monte Carlo simulations
and fully discussed by Rollettt \textit{et al}.\footnote{%
A. D. Rollett, D. J. Srolovitz, M. P. Anderson and R. D. Doherty, Acta Metall\textit{.}
40 (1992) 3475.
} Capillary forces, which induce a reduction of the total grain boundary energy,
can also be responsible for abnormal grain growth. In this case, normal grain
growth of the matrix grains is inhibited by either a texture effect or by the
presence of second phase particles. The former situation is well documented.
It is characterised by an anisotropy of the grain boundary energy and mobility,
as these last quantities depend on the relative misorientation of the grains.
Monte Carlo simulations have been performed on textured grain aggregates by
Rollett \textit{et al}.\footnote{%
A.D. Rollett, D.J. Srolovitz and M.P. Anderson, Acta Metall\textit{.} 37 (1989)
1227.
} and Grest \textit{et al}.\footnote{%
G. S. Grest, M. P. Anderson, D. J. Srolovitz and A. D Rollettt, Scripta Metall\textit{.}
24 (1990) 661.
} and were found in good agreement with the analytical approach proposed by W\"{o}rner
\textit{et al}.\footnote{%
C. H. W\"{o}rner, S. Romero and P. M. Hazzledine, J. Mater. Res. 6 (1991) 1773.
} However, situations are met in equiaxed polycrystals, where no texture is evidenced
either after a prestrain or after a further heat treatment, so that this mechanism
cannot be involved. 

The present study was motivated by the observation of abnormal grain growth
in several nickel base superalloys for high temperature applications produced
by powder metallurgy techniques.\footnote{%
M. Soucail, M. Marty and H. Octor, Scripta Metall. 34 (1996) 519.
}\( ^{,} \) \footnote{%
R. Messina, M. Soucail, T. Baudin and L.P. Kubin, J. Appl. Phys. 84 (1998) 6366.
} The microstructure of these alloys is made up principally of an austenitic
\( \gamma  \) \textit{}matrix, exhibiting no significant texture and hardened
by several families of L1\( _{2} \) ordered \( \gamma ' \) \textit{}precipitates.
The composition of these two phases is based on Ni\( _{3} \)Al, with significant
amounts of other elements like Co, Cr and Ti. Moderate or drastic abnormal grain
growth was found to occur during heat treatments performed above the \textit{}\( \gamma ' \)
solvus (typically at 1205\ensuremath{°}C). In such conditions, the only precipitates
that remain stable in the alloy are incoherent oxy-carbide particles, which
are preferentially located at the previously existing precipitate boundaries
with a volume fraction of about 0.1\%. These particles should effectively play
the role of obstacles to grain boundary migration, due to their small diameter
(about 100 nm) and spacing, and should normally impede normal as well as abnormal
grain growth. The models by Hillert\footnote{%
M. Hillert, Acta Met. 13 (1965) 227.
} and Andersen \textit{et al}.\footnote{%
I. Andersen, \O{}. Grong and N. Ryum, Acta Metall\textit{.} 43 (1995) 2689.
} refer to such a situation and are frequently invoked to explain the experimental
results. However, numerical simulations have not up to now been able to check
the predictions of these models.\footnote{%
D. J. Srolovitz, G. S. Grest and M. P. Anderson, Acta metall\textit{.} 33 (1985)
2233.
} 

Monte Carlo simulations were carried out in a model material consisting of an
equiaxed, fully recrystallized and isotropic grain structure containing a fine
dispersion of stable precipitates. A few useful definitions are briefly recalled
in Part 2, where, in addition, the bases of the Monte Carlo simulation are outlined.
The numerical results obtained in the absence or presence of thermally activated
unpinning of grain boundaries from precipitates are reported in part 3. These
results are discussed in part 4.

\section{Numerical model}

The numerical procedure used in the present work is similar to the one employed
in previous studies.\footnote{%
M. P. Anderson, D. J. Srolovitz, G. S. Grest and P. S. Sahni, Acta Metall. 32
(1984) 783.
}\( ^{,} \)\footnote{%
D. J. Srolovitz, M. P. Anderson, G. S. Grest and P. S. Sahni, Acta Metall. 32
(1984) 1429.
} It is described in detail elsewhere.\footnote{%
M. Soucail, R. Messina, A. Cosnuau and L. P. Kubin, Mater. Sci. Eng. A 271 (1999)
1. 
}\( ^{,} \) \footnote{%
R. Messina, Doctoral thesis, University Paris-Sud, (1998) Orsay.
} Use is made of a two-dimensional triangular lattice, with periodic boundary
conditions. To each lattice site \textit{S}\( _{i} \) is associated a crystallographic
orientation, represented by a number \textit{Q} ranging between 1 and 1000.
A grain is defined as an ensemble of adjacent sites having same orientation.
The initial microstructure is obtained by distributing at random the crystallographic
orientations on the lattice sites, which results in an initial microstructure
containing grains of uniform size and of dimension one lattice site. 

The particles are represented by fixed sites having a particular orientation,
\textit{S}\( _{i} \)\textit{= 0}. In two dimensions, and for incoherent particles,
the pinning force is entirely determined by the line tension of the grain boundaries.\footnote{%
E. Nes, N. Ryum and O. Hunderi, Acta Metall. 33 (1985) 11.
} Thus, the interfacial particle-matrix energy is assimilated the grain boundary
energy per unit length \textit{J}. In the initial microstructure the particles
are distributed at random with a surface fraction \textit{f} whose value is
0.5, 1 or 2\%. In what follows, we focus on the simulation results obtained
with \textit{f} = 2\%. 

The Hamiltonian \textit{H} of the system is written 

\begin{equation}
\label{eq.H}
H=\frac{{1}}{{2}}J\sum _{i=1}^{N}{\sum _{j=1}^{nn}{\left( {1-\delta _{S_{i}S_{j}}}\right) }},
\end{equation}

\noindent where \( \delta  \) is the Kronecker symbol and \textit{J} is a constant,
assuming that the microstructure is made up of an equiaxed and non-textured
polycrystal containing large-angle boundaries with isotropic energy. The sum
is taken over all nearest neighbours (\textit{nn}). The evolution of the microstructure
is simulated with the help of a Monte Carlo scheme. A site is chosen at random
and its orientation is randomly changed to one of the other (\textit{Q-}1) orientations.
If the resulting total variation in energy, \textit{\( \Delta  \)H}, is negative
or null, the reorientation is accepted. If the change in energy is positive,
the reorientation is accepted with the Boltzmann probability 

\begin{equation}
\label{eq.Boltzmann}
P=exp\left( {-\frac{{\Delta H}}{{kT}}}\right) 
\end{equation}

As the Hamiltonian only includes the grain boundary energy, the only driving
force for grain growth included in the simulation is a capillary force. The
time scale of the simulation is assimilated to the Monte Carlo Step (MCS), i.e.,
to the time corresponding to one reorientation attempt on all the sites that
do not contain a particle. A first set of simulations, performed at the absolute
zero of temperature, will be presented in the next section. The case of non-zero
temperatures is also considered, since the thermally activated unpinning of
grain boundaries from particles is a plausible event at high temperature. 

According to the two dimensional models of Hillert\( ^{7} \) and Andersen \textit{et
al}.\( ^{8} \), a necessary condition for obtaining abnormal grain growth is
the pre-existence of grains of size larger than the average size of the matrix
grains. We define by \( D_{ab} \), \( \overline{D} \) and \( \overline{D}_{lim} \)
the initial diameter of the future abnormal grain, the current mean diameter
of the matrix grains and the stagnant (i.e., final) value of \( \overline{D} \),
respectively. The results of the models can then be expressed in terms of two
reduced parameters, the diametrical size advantage, \( D_{ab}/\overline{D} \),
and \( \overline{D}/\overline{D}_{lim} \) the degree of matrix pinning. The
criteria for abnormal grain growth are written as follows : 

\begin{equation}
\label{eq.AGG-criterion-a}
\frac{{dD_{ab}}}{{dt}}>0
\end{equation}

\noindent and 

\begin{equation}
\label{eq.AGG-criterion-b}
\frac{{d}}{{dt}}\frac{{D_{ab}}}{{\overline{D}}}>0.
\end{equation}

Eq. (\ref{eq.AGG-criterion-a}) simply expresses that the large grain must grow,
while Eq. (\ref{eq.AGG-criterion-b}) expresses the condition that it must grow
faster than the matrix grains. This last condition destroys the self-similarity
of the coarsening process, which characterises normal grain growth. However,
As was shown numerically by Srolovitz \textit{et al}.\( ^{9} \), the pre-existence
of large grains is not a sufficient condition to obtain abnormal grain growth
and one must consider, in addition, the influence of the particles on the growth
process. Their effect is essentially to reduce the local curvature of the matrix
grains and, therefore, their capillary driving force. As a consequence, a large
grain may, then, be submitted to a driving force larger than that of the matrix
grains and grow at a sufficient rate to fulfil the second criterion for abnormal
growth {[}Eq. (\ref{eq.AGG-criterion-b}){]}. 

In the present simulations, a large test grain with a pseudo-circular shape
and an adjustable diameter is initially introduced into a matrix structure.
The latter may have an average initial grain size \( \bar{D}_{o} \) of one
site, or may have been previously subject to a sequence of uniform growth in
order to increase its initial degree of pinning. The initial value of the size
advantage,\( \left( {D_{ab}/\bar{D}}\right) _{o} \) is fixed to a value of
about 3, except if stated otherwise. This initial value is implemented within
a minor error due to the discrete nature of the simulation. Because the growing
grains sometimes exhibit irregular shapes, it is convenient to monitor their
growth by measuring their area. As a consequence, we make use in what follows
of a reduced area advantage, \( A_{ab}/\overline{A} \), and a reduced area
degree of pinning, \( \overline{A}/\overline{A}_{lim} \). These two quantities
are obviously identical to the equivalent diameters squared.

\section{Results}

\subsection{\textit{Mechanical approach (T = 0 K)}}

In a first step, we consider a pure mechanical approach (T = 0 K) and investigate
the role of the pinning degree of the matrix grains on the coarsening process
of a large test grain. The stagnant matrix grain size is known from a previous
simulation study\( ^{12} \) on Zener pinning, i.e., on the pinning of uniformly
growing grains in the presence of precipitates.\footnote{%
C. S. Smith, Trans. metall. Soc. A.I.M.E\textit{.} 175 (1948) 15.
} Fig. 1(a) shows the evolution with time of the mean grain size of a polycrystal
containing a surface fraction \textit{f} = 2\% of particles. The driving force
for grain growth in the presence of particles is provided by the grain boundary
curvature, which decreases with increasing time and mean grain size. A steady
pinned configuration is obtained after a certain amount of time, when the driving
force becomes insufficient to overcome the Zener drag. As was shown recently,
such a steady state can also be obtained by a deterministic approach.\footnote{%
D. Weygand, Y Br\'{e}chet and J. L\'{e}pinoux, Acta Metall. 47 (1999) 961.
} The stagnant microstructure corresponding to the conditions of Fig. 1(a) is
represented in Fig. 1(b). 

A potentially anomalous grain with an initial reduced size advantage \( (D_{ab}/\bar{D})_{0}\approx 3 \)
is then inserted into matrix structures with various degrees of pinning. The
evolution with time of the area \textit{A}\( _{ab} \) of the test grain and
of the reduced area advantage, \( A_{ab}/\overline{A} \), are shown in Figs.
2(a) and (b), respectively. Here and in what follows, the surface fraction of
particles is \textit{f} = 2\%. 

The ratio \( A_{ab}/\overline{A} \) increases monotonically with time and further
saturates, provided that the initial degree of pinning is sufficiently large.
Hence, the test grain can grow abnormally only if the matrix grains are not
too far from their limit size. When the initial degree of pinning is small and
the matrix grains are highly mobile, no abnormal grain growth occurs because
the size advantage of the large grain gradually vanishes during the growth process.
Figure 3 shows a sequence of microstructural evolution leading to abnormal grain
growth. One can notice that the matrix has not evolved significantly in comparison
to the large grain. 

The initial pinning degree of the matrix grains is now fixed to a value (\( \overline{D}/\overline{D}_{lim} \))\( _{o} \)
and the initial size advantage of the large grain is varied. Figs. 4(a) and
(b) show the evolution of the area A\( _{ab} \) and of the reduced area advantage,
\( A_{ab}/\overline{A} \), of the large grain for different values of the initial
size advantage. The large grain grows when the size advantage is beyond a certain
critical value and shrinks otherwise {[}cf. Fig. 4(a){]}. The growth is abnormal
in the sense of equation 4 when the initial size is above a second critical
value {[}cf. Fig. 4(b){]}, which is larger than the first one (cf. Fig. 6, below).
This confirms an assumption of the analytical models,\( ^{7,8} \) according
to which the pre-existence of a large grain is necessary for abnormal grain
growth to occur. 

The size of the test grain saturates after a certain amount of time and the
whole microstructure is then immobilised by Zener pinning. The large grain becomes
pinned after the matrix grains, since it has a larger driving force governing
its growth. Fig. 5 shows a sequence of microstructural evolution starting with
an initially pinned microstructure, \( \overline{D}_{o}=\overline{D}_{lim} \),
and an initial size advantage \( \left( {D_{ab}/\bar{D}}\right) _{o}=6 \).
This situation approximately corresponds to the one studied by Srolovitz et
al.,\( ^{9} \) where no abnormal grain growth was observed. In Fig. 3, one
can see that several unpinning events have occurred on the boundary of the test
grain before it became fully stabilised. In contrast, no unpinning occurs in
the sequence of Fig. 5, although the size advantage is larger. The boundary
of the abnormal grain moves until it reaches the nearest particles. Then, it
adopts a minimum energy configuration with straight segments of boundaries between
the pinning particles. 

The previous results are synthesised in Fig. 6, which shows a map of the different
types of behaviour obtained, viz. abnormal grain growth, grain growth but no
abnormal growth and grain shrinkage, as a function of the initial values of
the reduced size advantage and pinning degree. This diagram confirms that for
abnormal grain to occur, the initial relative size (\( D_{ab}/\overline{D} \))\( _{o} \)
must be within a certain range of values whose upper bound is about 4. This
suggests that the condition of Eq. (\ref{eq.AGG-criterion-b}), according to
which the growth rate of the large grain must be larger than that of the matrix
grains, is more difficult to fulfil when the initial size advantage is too large.
The reason can be better understood by expressing the second condition for abnormal
grain growth {[}Eq. (\ref{eq.AGG-criterion-b}){]} as follows : 

\begin{equation}
\label{eq5}
{{\frac{{dD_{ab}}}{{dt}}}\mathord {\left/ {\vphantom {{\frac{{dD_{ab}}}{{dt}}}{\frac{{d\overline{D}}}{{dt}}}}}\right. \kern -\nulldelimiterspace }{\frac{{d\overline{D}}}{{dt}}}}>\frac{{D_{ab}}}{{\overline{D}}}.
\end{equation}

One can see from Eq. (5) that with a larger initial size advantage, a larger
growth rate of the test grain is required to fulfil the criterion. Srolovitz
\textit{et al}.\( ^{9} \) concluded from their simulations that abnormal grain
growth is not possible in equiaxed materials containing particles when the only
driving force is of capillary nature. However, this conclusion was reached from
a study involving too large size advantages and we see from the present results
that an intermediate domain can be defined where abnormal grain growth is, indeed,
possible. 

Finally, the type of abnormal grain growth that has been obtained so far can
be considered as moderate, but certainly not as drastic (cf. for instance Fig.
3). The objective of the next sections is to understand in which conditions
the present simulation can reproduce the experimental observations of drastic
abnormal growth.

\subsection{\textit{Thermal unpinning}}

As drastic grain growth is experimentally a high temperature feature, it may
be necessary to account for the possibility of a thermally activated detachment
of the grains from the pinning particles. Such an effect was already studied
analytically by Gore \textit{et al.}\footnote{%
M. J. Gore, M. Grujicic, G. B. Olson and M. Cohen, Acta Metall. 37 (1989) 2849.
} and with the help of extensive Monte Carlo simulations by Miodownick \textit{et
al.},\footnote{%
M. Miodownik, E. A. Holm and G. N. Hassold, Scripta Metall. 42 (2000) 1173.
} in the case of \textit{homogeneous} grain growth in the presence of pinning
particles. 

In our simulation, such a mechanism can be achieved by allowing site re-orientations,
which involve the unpinning of a grain boundary and lead to an increase of the
energy of the system, to be conditionally accepted with a Boltzmann probability.
Thus, in this first step, it is simply assumed that the reduced energy \textit{J/kT}
is not too large, so that the unpinning probability is significant. Site reorientations
involving no pinning events are treated in a purely mechanical manner (\textit{T}
= 0 K). The initial microstructure consists of structure of pinned grains, like
the one of Fig. 2. A single test grain test is introduced in the pinned microstructure
and from the beginning of the simulation, the temperature is set to a non-zero
value. 

Fig. 7 shows the evolution of a large test grain as a function of the reduced
temperature. Abnormal grain growth is never obtained within the range of reduced
temperatures considered here, 0.2 \textit{J/k} < \textit{T} < 0.6 \textit{J/k}.
In addition, an apparently paradoxical behaviour is obtained. The higher the
temperature, the earlier the size advantage, \( {{D_{ab}}\mathord {\left/ {\vphantom {{D_{ab}}{\overline{{\kern 1pt}D}}}}\right. \kern -\nulldelimiterspace }{\overline{{\kern 1pt}D}}} \),
starts decreasing. Indeed, one may expect \textit{a priori} that the size advantage
should increase with time, all the more as temperature increases. The interpretation
of this particular behaviour is postponed to the discussion part. Finally, Fig.
8 shows a typical microstructural evolution at a temperature \textit{T} = 0.4
\textit{J/k}. Similar results, not shown here, were obtained with other values
of the particle volume fraction, 0.5\% and 1\%, as well as with other values
of the particle size (3 and 7 sites).\( ^{13} \) As will be discussed in the
next section, the reason for the absence of abnormal growth stems from the too
rough treatment of the unpinning events.

\subsection{\textit{Selective thermal unpinning }}

The discrete simulation attributes the same activation energy for unpinning
to the small and the large grains. However, it is important to have in mind
in mind, that in conditions favourable for abnormal grain growth, the small
grains are subject to a smaller capillary driving force than the large ones.
Hence the corresponding activation energy should be increased accordingly. As
the present simulation is not refined enough to describe the local configurations
of the pinned grain boundaries and include this effect, the difference in energy
barrier between the large and small grains is incorporated into the Metropolis
criterion in a very simplified manner. The method employed is illustrated by
Fig. 9. Between two pinning events, the boundaries of the test grain evolve
according to the capillary force i.e., the local curvature, and its motion is
not thermally activated. When the boundary is pinned by a particle, the probability
for thermal unpinning is considered. For the small grains, the probability for
thermal unpinning is assumed negligible and only mechanical unpinning can occur.
This rather crude procedure may seem to drive the final result but, as discussed
in the next part, it can be justified in physical terms. 

Fig. 10 shows the evolution of the reduced area \( {{A_{ab}}\mathord {\left/ {\vphantom {{A_{ab}}{\overline{A}}}}\right. \kern -\nulldelimiterspace }{\overline{A}}} \)
of the abnormal grain as a function of time and for different reduced temperatures.
Drastic abnormal grain growth does occur above a critical temperature \textit{T}\( _{c} \)
\textit{\( \approx  \)} 0.2 \textit{J/k} and qualitatively follows an Arrhenius
type of behaviour. Fig. 11 shows the microstructural evolution at \textit{T}
= 0.3 \textit{J/k} of a polycrystal containing a large circular grain, initially
introduced in a pinned matrix structure. In such conditions, an initial size
advantage of about 3 is sufficient for the large grain to invade the whole microstructure.
Similar results, not reported here, were obtained with other values of the particle
surface fraction (0.5\% and 1\%). It was also checked that such a strong abnormal
grain growth occurred with two other particle sizes, 3 and 7 sites. This type
of behaviour strongly contrasts with the moderate growth of the large grain
obtained in the absence of selective unpinning. Below \textit{T} = 0.2 \textit{J/k},
this drastic abnormal growth behaves as a transient process. The whole microstructure,
including the large grain, becomes eventually pinned when the driving force
is no longer sufficient to overcome the Zener drag.

\section{Discussion}

In the present study, the pre-existence of a large grain has systematically
been assumed and an initial size advantage was confirmed to be a necessary condition
for obtaining abnormal grain growth. The presence of these large grains can
be explained as follows. The thermo-mechanical treatment induces a weak primary
recrystallization, thus generating a few fresh grain embryos. The latter may
further grow at the expense of the deformed matrix grains. For this process
to occur, it is sufficient to slightly predeform the material close to a critical
deformation value.\( ^{13,} \)\footnote{%
C. Antonione, F. Marino, G. riontino, and M. C. Tabasso, J. Mater. Sci\textit{.}
12 (1977) 747.
} The driving force for this primary recrystallization step is, then, provided
by the stored energy accumulated during the prestrain. When this driving force
is exhausted, the coarsening process can only continue under the influence of
the capillary driving force. 

A second condition for abnormal grain growth is that the mobility of the matrix
grain boundaries must be sufficiently reduced by pinning effects. This allows
the large grain to outgrow the surrounding matrix grains. When the matrix grain
boundaries are too mobile, i.e. when \( \overline{D}/\overline{D}_{lim} \)
is small, the large grain may grow but not in an abnormal manner. This situation
is similar to the one encountered in the absence of precipitation, when a large
grain is introduced in a matrix governed by normal grain growth. It has been
shown both analytically\footnote{%
C. V. Thompson, H. J. Frost and F. Saepen, Acta metall\textit{.} 35 (1987) 887.
} and numerically,\( ^{9} \) that such a large grain becomes rapidly incorporated
into the steady-state distribution function characterising normal grain growth.
Therefore, a significant initial perturbation is necessary in order to destabilise
the uniform growth process. 

The map of coarsening behaviour shown in Fig. 6 agrees well, in qualitative
terms, with the predictions of the model by Andersen \textit{et al}.\( ^{8} \)
In both cases, an upper limit is predicted for the initial size advantage leading
to abnormal grain growth. Such a feature is absent from the classical model
of Hillert.\( ^{7} \) Indeed, Hillert's approach is a pure mean-field model,
whereas the approach by Andersen et al.\( ^{8} \) combines a mean-field aspect
with a local aspect. The latter is associated with a fine description of the
geometry of the interface between the large grain and the surrounding matrix
grains. The mean-field aspect resides in the assimilation of the matrix grain
structure to a regular honeycomb structure, as first proposed by Gladman.\footnote{%
T. Gladman, Proc. R. Soc\textit{. A} 294 (1966) 298.
} 

It is interesting to compare the present results with those obtained in the
absence of particles with \textit{anisotropic} materials, i.e., with anisotropic
grain boundary energy and/or mobility. In the case of a mobility advantage and
a constant grain boundary energy, simulations\( ^{2,3} \) as well as the model
by W\"{o}rner \textit{et al}.\( ^{4} \) show that abnormal grain growth also
occurs within a range of values of the initial size advantage. Quite generally,
a factor has to inhibit normal grain growth, which can be provided by local
pinning, textural effects or surface energy anisotropy. 

As drastic abnormal coarsening is not obtained within a pure mechanical approach,
an attempt was made in section 3.2 to account for thermal effects. The physical
idea consists in taking into account thermal detachment and noting that this
later should be easier when the driving force for grain boundary motion is larger.
This was simply performed by treating the unpinning of the grain boundaries
from the second phase particles via the Metropolis criterion. As a result, it
was found that the tendency to abnormal grain growth was not enhanced but reduced,
even at elevated reduced temperatures (cf. Fig. 8). The obvious reason for this
unphysical behaviour is that the abnormal grain and the matrix grains are able
to pinch off the particles with the \textit{same} probability, since the activation
energy is not indexed on the local driving force due to the discrete lattice
structure. As a result, the advantage provided by the particles to the large
grain, due to the reduction of the average curvature radius of the small grains,
is destroyed. Indeed, in a more realistic (continuous) approach, the thermally
activated unpinning of the grain boundaries should be easier for the abnormal
grains, due to their larger mean curvature, than for the small matrix grains.\( ^{17,21,} \)\footnote{%
C. H. W\"{o}rner and A. Olguin, Scripta Metall. 28 (1992) 1.
} A similar behavior was observed for all particle sizes (1, 3 and 7 sites).
This is in agreement with theory since, in 2D, the maximal pinning force is
written \( F_{pin}=2\gamma _{l} \), where \( \gamma _{l} \) is the grain boundary
line tension, and does not depend on the particle size.\( ^{14} \) In contrast,
one has \( F_{pin}\propto \gamma _{s}r \) in 3D, where \( \gamma _{s} \) is
the grain boundary surface tension and \textit{r} is the average radius of the
particles.\( ^{14} \) This explains why in recent 3D simulations the pinning
and unpinning properties of a single grain boundary were found to strongly depend
on the particle size.\footnote{%
M. Miodownik, J. W. Martin and A. Cerezo, Phil. Mag. A79 (1999), 203.
} 

The present procedure leading to drastic abnormal grain growth may seem a bit
rough. Restricting thermal activation to the pinning sites that involve the
boundaries of the large grain indeed favours the growth of the latter. However,
although this method may be considered as an \textit{ad hoc} one, it is not
devoid of physical meaning. It simply consists of replacing a continuum energy
spectrum that is distributed over two rather different populations by two average
characteristic states. It is possible to verify that this method does not oversimplify
too much the dependence of the energy barrier on the driving force. For this
purpose, one specific grain of the pinned matrix was selected and allowed to
thermally pinch off the particles. As can be seen from Fig. 12, this grain does
not grow faster than the surrounding matrix at a temperature \textit{T} = 0.3\textit{J/k},
while in the same conditions the dimension of the large grain significantly
increases (cf. Fig. 11). This result shows that the dependence of the energy
barrier on the driving force of a specific grain is well reproduced in a qualitative
manner. It also illustrates in a qualitative way how the coupled effect of a
critical size and non-zero temperature is needed to trigger true anomalous grain
growth. 

W\"{o}rner and Olguin\( ^{22} \) have shown that thermal unpinning becomes
significant for particle sizes smaller than about 100 nm, low particle volume
fractions (smaller than 0.3\%) and sufficiently high annealing temperatures
(about \textit{}1000\ensuremath{°}C). These values are quite consistent with
the experimental ones in the case of a nickel-based superalloy {[}5{]} (cf.
part 1). In such conditions and with a grain boundary energy of typically 1.5
J/m\( ^{2} \), the energy barrier is estimated to about 0.6 eV for the abnormal
grain. The ratio \textit{kT/J} should then be about 0.2, which is within the
range investigated in the present study. Experimentally, abnormal growth is
obtained after annealing at 1205 \ensuremath{°}C for 4 hours. For longer annealing
times, the drastic coarsening process is such that one single grain invades
the whole microstructure, as is also the case in the present simulation.

\section{Conclusion}

The simulation of abnormal grain growth in polycrystalline aggregates has yielded
the main following results, that are qualitatively independent of the particle
surface fraction within the range investigated in the present study. 

1. The moderate growth of a test grain under the effect of capillary forces
can be effective in polycrystalline materials with equiaxed grains containing
stable particles. 

2. The simulations confirm that two interdependent reduced parameters, the reduced
abnormal grain size and the reduced matrix mean grain size, govern the occurrence
of this growth process. 

3. For moderate grain growth to occur, the initial size advantage of the \char`\"{}abnormal\char`\"{}
grain has to lye between an upper and a lower bound. In addition, the matrix
grain boundaries must be sufficiently pinned by second phase particles. These
results are in good agreement with the recent model by Andersen \textit{et al}.\( ^{8} \) 

4. True abnormal grain growth in an equiaxed and isotropic matrix containing
stable particles can be described as resulting from a thermally activated unpinning
process. It is still conditioned by an initial size advantage and is enhanced
by the effect of the capillary driving force on the large grain. 
\newpage

\textbf{FIGURE CAPTIONS} \\ \\

\underbar{FIG. 1.} a) - Evolution of the mean grain area \( \bar{A} \) (number
of sites) with Monte Carlo Step number, for a microstructure containing an area
fraction of particles \textit{f} = 2\%. 

b) - The corresponding pinned microstructure. \\ 

\underbar{FIG. 2.} Evolution with time of a test grain, for a fixed size advantage
\( \left( {A_{ab}/\overline{A}}\right) _{0}=\left( {D_{ab}/\overline{D}}\right) _{0}^{2}\approx 10 \),
and for different values of the initial degree of matrix pinning, \( \overline{D}_{0}/\overline{D}_{lim} \),
given in the insert. a) - Absolute area, \textit{A}\( _{ab} \). \textbf{}b)
- Reduced area, \( A_{ab}/\overline{A} \). \\

\underbar{FIG. 3}. A typical microstructure exhibiting abnormal grain growth.
The initial conditions are \( \left( {D_{ab}/\overline{D}}\right) _{0}=3 \)
and \( \overline{D}_{0}/\overline{D}_{lim}=0.9 \). \\ 

\underbar{FIG. 4}. Evolution of the area of the test grain for a fixed degree
of pinning of the matrix grains, \( \overline{D}_{0}/\overline{D}_{lim}=0.9 \)
and for various values of the relative size of the test grain, \( D_{ab}/\overline{D}_{0} \),
given in the insert. a) - Absolute area, \textit{A}\( _{ab} \). \textbf{}b)
- Reduced area, \( A_{ab}/\overline{A} \). \\

\underbar{FIG. 5.} Microstructural evolution of a large grain (with initial
size advantage \( \left( {D_{ab}/\overline{D}}\right) _{0}=6 \)) embedded in
an initially pinned polycristal (\( \overline{D}_{0}/\overline{D}_{lim}=1 \)).
No abnormal grain growth occurs. \\

\underbar{FIG. 6}. Map of the behaviour \textbf{}of the test grain in a plot
of size advantage \( \left( {D_{ab}/\overline{D}}\right) _{0} \) vs. pinning
degree (\( \overline{D}_{0}/\overline{D}_{lim} \)). The three types of behaviour
observed are abnormal grain growth (AGG), grain growth (GG) and grain shrinkage
(GS).\\

\underbar{FIG. 7.} Thermal unpinning. Evolution of the reduced area advantage,
\( {{\overline{A}_{ab}}\mathord {\left/ {\vphantom {{\overline{A}_{ab}}{\overline{A}}}}\right. \kern -\nulldelimiterspace }{\overline{A}}} \),
for different values of the temperature \textit{T}, expressed in units of \textit{J/k.}
The initial size advantage of the large grain is about 3. \\

\underbar{FIG. 8.} Microstructural evolution in conditions where thermal unpinning
of the grain boundaries from the precipitates is effective (\textit{T =} 0.4
\textit{J/k})\textit{.} The initial size advantage of the large grain (in the
centre of the simulation) is about 3. No abnormal grain growth is observed.\\

\underbar{FIG. 9.} Schematic procedure for simulating the thermally activated
unpinning of the boundaries of the abnormal grain. Three types of sites are
particularised. Sites P contain a particle, sites L belong to the large abnormal
grain and site S to a small grain. A Boltzmann reorientation is accepted only
if the considered site belongs to a small grain and is simultaneously adjacent
to a particle and to the abnormal grain. \\

\underbar{FIG. 10}. Evolution of the relative size, \( A_{ab}/\overline{A} \),
of the abnormal grain for different temperatures. Thermally activated unpinning
is restricted to the boundaries of the abnormal grain. The initial size advantage
of this grain is 3.\\

\underbar{FIG. 11}. Microstructural evolution in the case of selective thermal
unpinning (\textit{T =} 0.3 \textit{J/k).} The initial size advantage of the
large grain is about 3. True abnormal grain growth is observed. \\

\underbar{FIG. 12.} Effect of the thermally activated grain boundary unpinning
on a particular matrix grain (in grey). The temperature is \textit{T =} 0.3
\textit{J/k} and the initial microstructure is similar to the one of Fig. 2.
No significant evolution of this grain is observed. \\
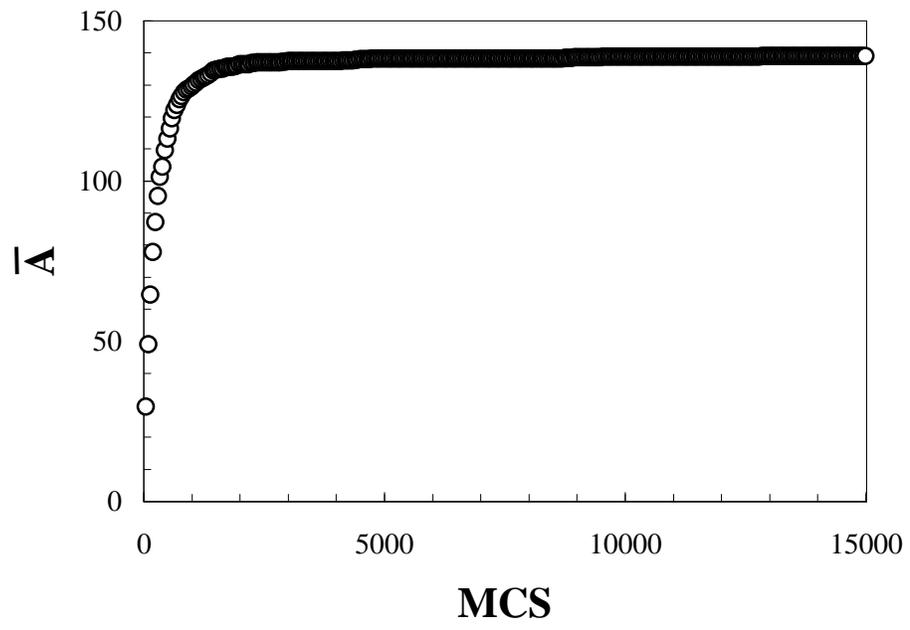
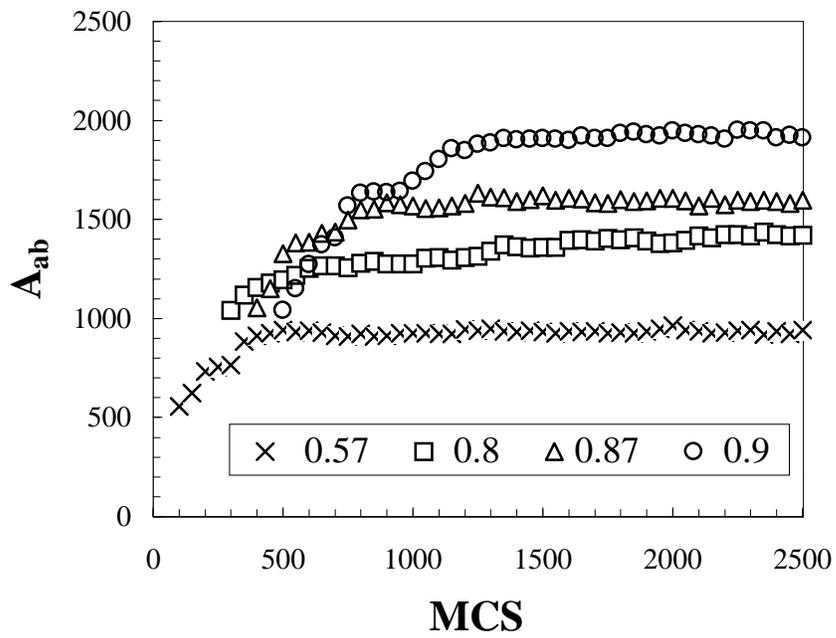
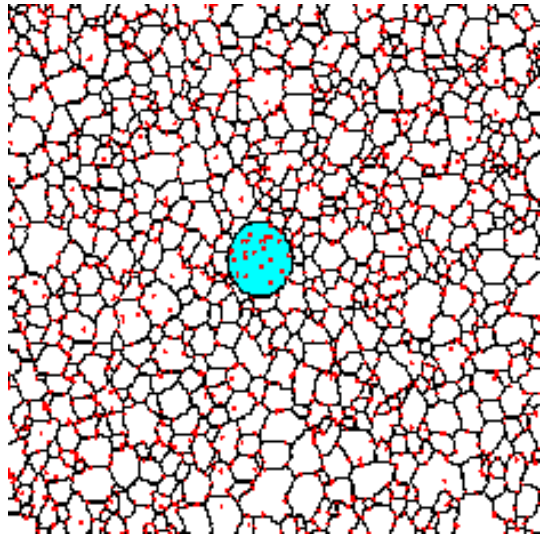
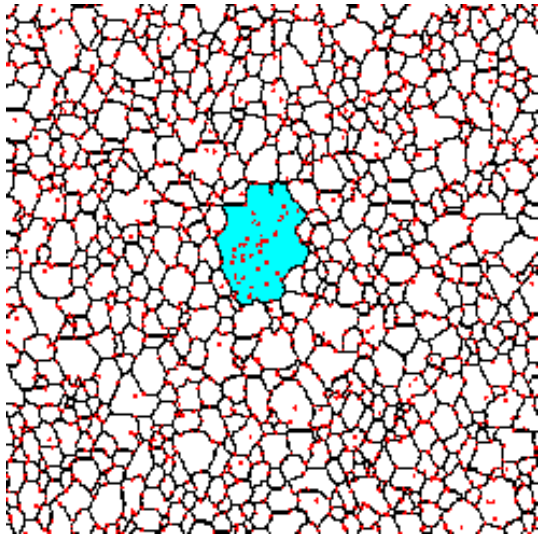
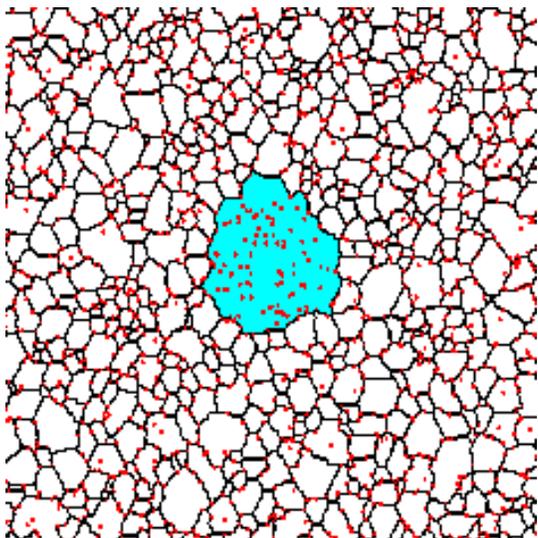
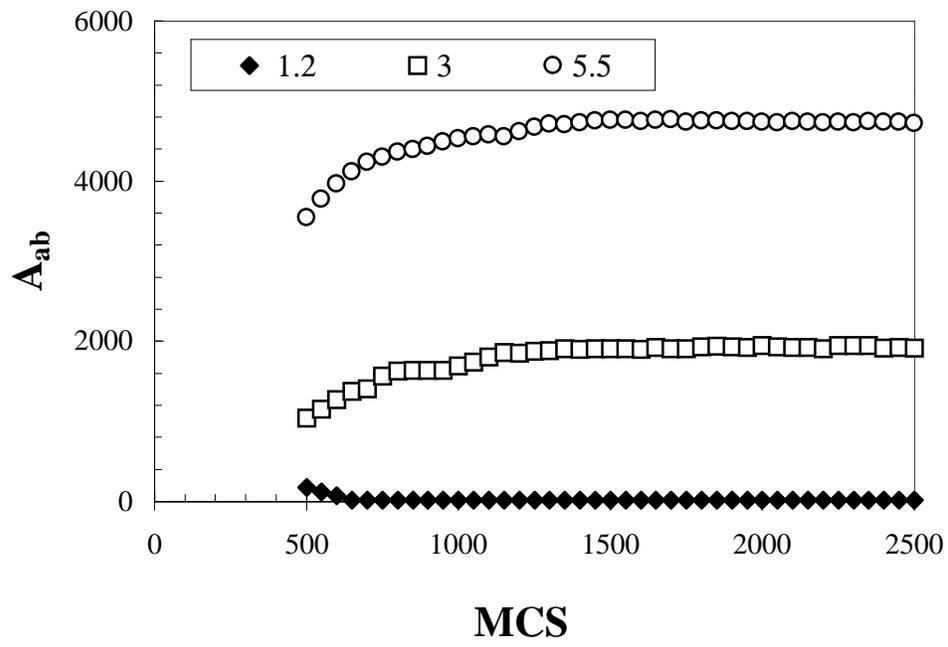
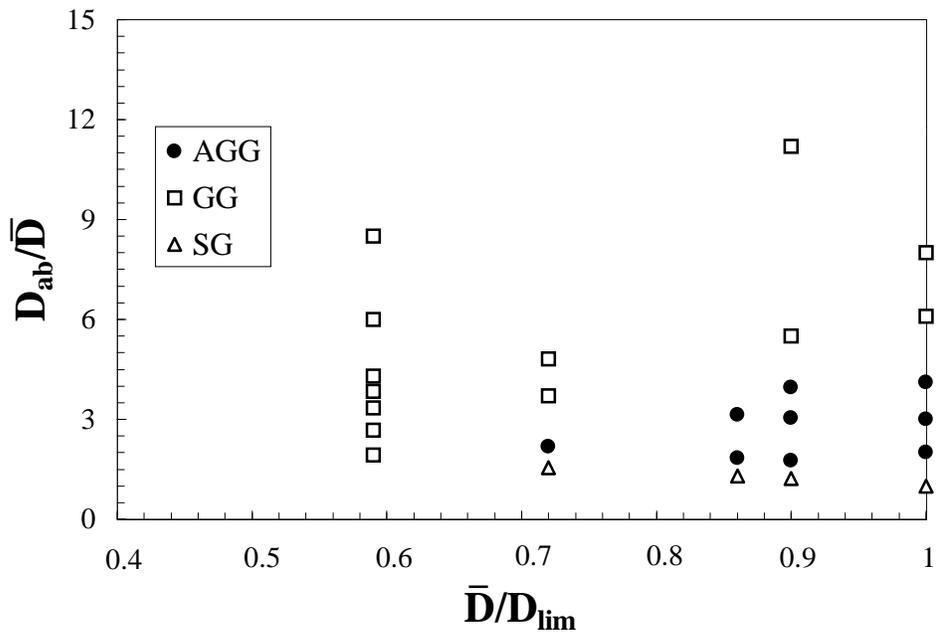
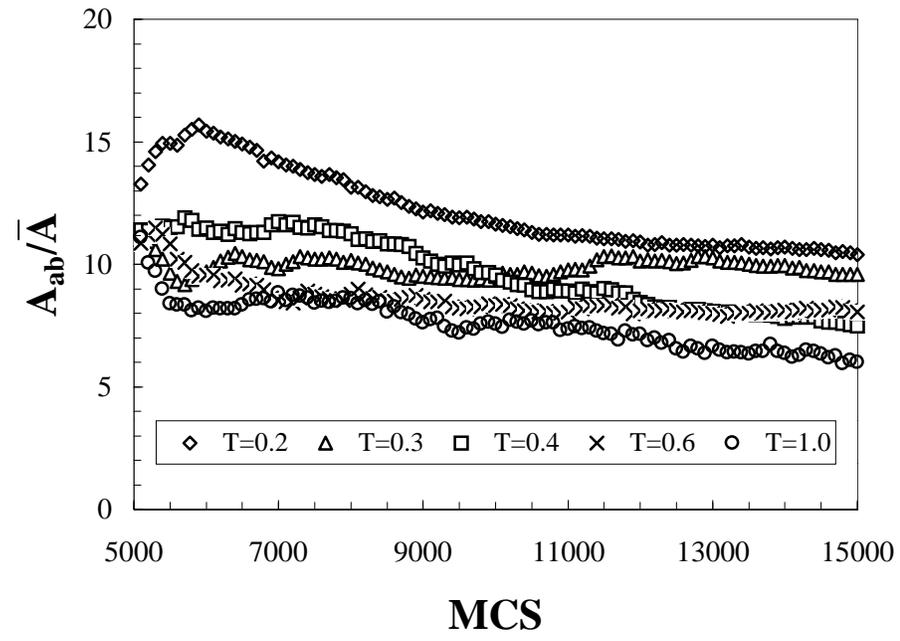
\end{document}